\documentclass[%
 reprint,
 amsmath,amssymb,
 aps,Physics,pre
]{revtex4-1}

\usepackage{graphicx}
\usepackage{dcolumn}
\usepackage{bm}
\usepackage{comment}
\usepackage{xcolor}

\begin{document}

\newcommand{\add}[1]{\color{black}{#1}}
\definecolor{comment}{RGB}{0,0,0}

\preprint{APS/123-QED}

\title{Machine learning and predicting the time dependent dynamics of \\ local yielding in dry foams}

\author{Leevi Viitanen}
\author{Jonatan R. Mac Intyre}
\author{Juha Koivisto}
\author{Antti Puisto}
\author{Mikko Alava}

\affiliation{%
Aalto University, School of Science, Department of Applied Physics, P.O.B 11100, 00076 Aalto, Finland
}%

\date{\today}


\begin{abstract}
The yielding of dry foams is enabled by small elementary yield events on the bubble scale, ``T1''s.
We study the large scale detection of these in an expanding 2D flow geometry using artificial intelligence (AI) and nearest neighbour analysis.
A good level of accuracy is reached by the AI approach using only a single frame, 
with the maximum score for vertex centered images highlighting the important role the vertices play in the local yielding of foams.
We study the predictability of T1s ahead of time and show that this is possible on a timescale related to the waiting time statistics of T1s in local neighborhoods.
The local T1 event predictability development is asymmetric in time, and measures the variation of the local property to yielding and similarly the existence of a relaxation timescale post local yielding.
\end{abstract}


\maketitle


\section{Introduction}

Dry foams are assemblies of gas pockets separated by thin films of liquid forming a connected polygonal film structure \cite{Weaire1999} at a configuration globally minimizing the surface energy~\cite{brakke1992}.
A finite external stress is needed to make foams flow and, due to the absence of thixotropy, foams are usually rheologically characterized as simple yield stress fluids~\cite{moller2009attempt}.
The steady-state, global flow curves of foams are considered to be typical examples of Herschel-Bulkley fluid-like behavior~\cite{ovarlez2013existence, barry2010shear}.

At bubble scale, the viscoplastic flow of dry foams is enabled by small elementary topological yield events referred to as T1's and T2's, analogous for instance to shear transformation zones (STZs) in amorphous solids \cite{Falk1998, ManningPRL2011}. 
The T2 events involve the disappearance of three-sided bubbles, while
the T1 events 
refer to a neighbour swap between four bubbles. Both events enable the system to jump from one metastable surface energy minimum to another, including a local relaxation
of the stored elastic energy of the foam~\cite{Petit2015, Cohen-Addad2013}.
T2 events require either film breakup or gas exchange to occur, making them less frequent compared to T1s in systems under continuous deformation.
A quasi-2D setting or simplifying geometry, is often used to study the flow of foams and T1s therein~\cite{Dollet2007, VanHecke2010, Chevalier2017, viitanen2019, viitanen2019biores}.
The T1 events can be either reversible or non-reversible depending on the geometric configuration and the stress direction~\cite{Lundberg2008, Keim2013}.

\begin{figure}[!t]
    \centering
    \includegraphics[width=0.8\columnwidth]{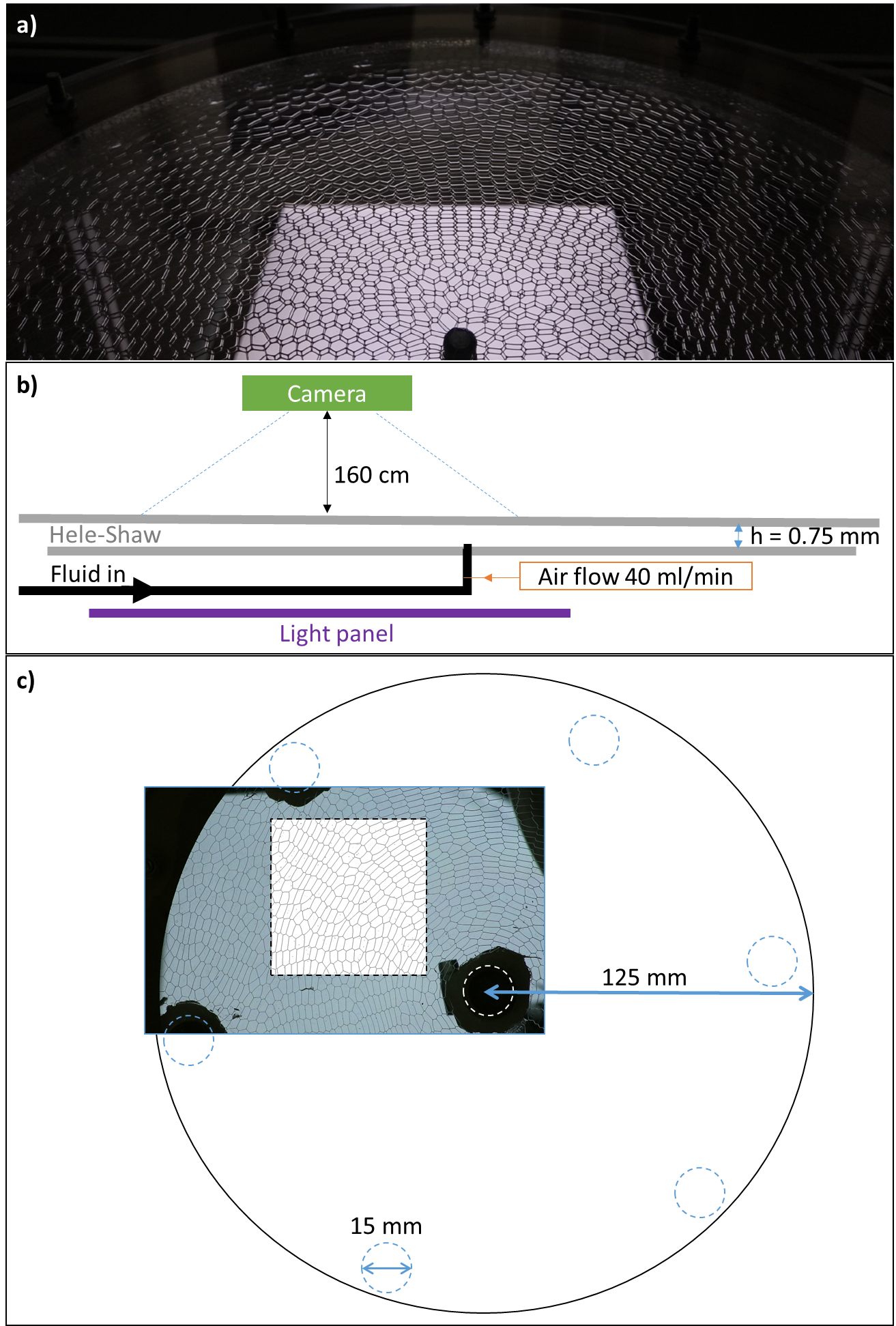}
    \caption{Radially symmetric 2D Hele-Shaw cell creates an expanding flow field.
    a) An angled photograph of the Hele-Shaw cell shows dry foam with typically hexagonal bubbles.
    The 15 mm inlet hose is located at the bottom center of the device. 
    b) Schematic side view illustrates how the fluid and air are foamed at the inlet.   
    c) The schematic drawing of the top view of the circular Hele-Shaw cell is overlaid by a
    single raw image from the video stream. The dashed line illustrates the analysis area that is skeletonized to a binary image. The inlet at the center and outlet holes have diameter $d_h = 15 $ mm shown as dashed blue circles.}
    \label{fig1:device}
\end{figure} 
In the macroscopic flow and yielding of foams the plastic deformation arises from a complex dynamics of T1 events with localization and clustering \cite{gopal1995nonlinear,Durian1997,Kader1999}.
Here, we study T1 events from large-scale statistics. We first detect tens of thousands of events from bubble raft flow dynamical data by comparing two sequential frames using a nearest neighbor detection algorithm \cite{Dollet2007}.
Second, we feed single frames to a Convolutional Neural Network (CNN) and successfully predict the T1 events by the film structure and extract the essential features.
We use this tool to explore T1 dynamics and the role of the film structure by varying the region of interest (ROI).
The relation of the T1 dynamics to macroscopic flow have been studied and interestingly, the time dependent deformation tensor \cite{Dollet2010,Marmottant2008} and 
the strain rate \cite{Wang2007}, both dependent on the local velocity, were found to correlate with the T1 frequency while it has been suggested there is no correlation between the snapshot of the film structure and the T1 frequency~\cite{Dollet2007}.
This would imply T1 detection based on a single structure snapshot should fail most of the time.

We find that T1 events can be predicted by the localized changes in film vertices, while the bubble shape itself does not contain the same information. 
The local predictability evolution is shown by time-dependent AI predictions to be asymmetric in time, and we interpret the results as a measure of the decrease in local yield stresses and as a manifestation
of the inherent local relaxation after a T1. 
The foam structure exhibits weaker and stronger spots, be it due to the network structure or film properties~\cite{Denkov2009}.
In amorphous solids soft spots have been studied using measures for non-affine deformation \cite{SchoenholzPRX2014}, 
Voronoi cell anisotropy \cite{RieserPRL2016}, and machine learning tools \cite{CubukPRL2015, SharpPNAS2018} and the distribution of local yield stresses
is an inherent property of plasticity models \cite{Nicolas2018RMP} as is the local relaxation dynamics after a yield event 
\cite{Rodney2011}.

We compare two methods of identifying T1 events from a video of a two-dimensional flow. 
First, we detect tens of thousands of events from dynamical data by comparing two sequential frames using well-established detection methods.
Second, we feed single frames to neural network and successfully predict the T1 events by the snapshot of the film structure. 
Our main findings show that T1 events are correlated with the localized change in film vertices, 
while the bubble perimeter itself does not contain similar information.

\section{Methods}  

Fig.~\ref{fig1:device} illustrates the circular Hele-Shaw cell with a bubble inlet at the center used in the experiments. The flow created is symmetric with a radial expansion that ensures T1 events will be present due to shear \cite{Dollet2015}. {\color{comment}In addition, due to the decreasing flow velocity towards the edges of the cell, a single experiment inherently probes the system at a range of shear rates.}
The cell diameter is $d = 250$ mm and the gap height is $h = 0.75$ mm.
The inlet has the diameter of $15$ mm. 
Foaming is facilitated by reducing the water surface tension by mixing common household
dishwashing liquid (Fairy) with the weight fraction of 1 to 39. %
The solution viscosity is indistinguishable from that of water, $\eta = 1~\mathrm{mPa}\cdot\mathrm{s}$,
 confirmed using Anton Paar 302 rheometer in a bob-and-cup geometry in the shear rate range of $\dot{\gamma} = 0.1$ 1/s to $\dot{\gamma} = 100$ 1/s.

\begin{figure}
    \centering
    \includegraphics[width=3.4in]{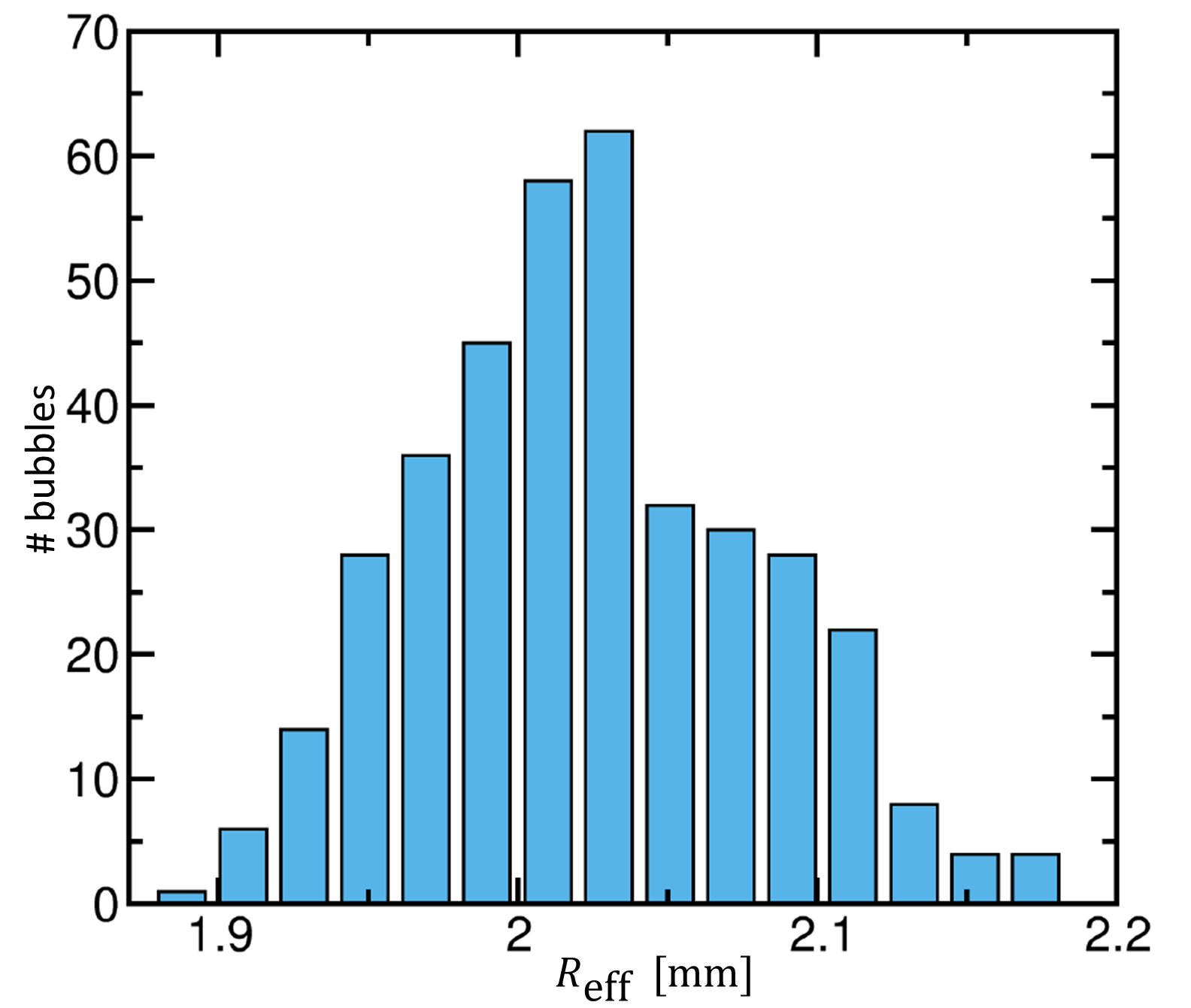}
    \caption{The foam studied is rather mono disperse having the average effective radius 
    $R_{eff} = 2.0\pm 0.1$ mm, where the error is the standard deviation.
    The histogram contains bubble sizes from three snapshots separated by 200 s with data from 378 bubbles.}
    \label{fig:bubble_properties}
\end{figure}

Both, the fluid and air, are driven into the cell using a constant pressure set by two SMC ITV0010 pressure controllers and a manual flow limiter.
The inlet flow rates are set to give an average bubble velocity of $v=1.0$ mm/s.
The air is injected through a 26-gauge needle.
The outer rim outlets are open to the ambient pressure.
The imaging is implemented using a Canon M3 camera, which takes compressed video at $1920 \times 1080$ pixels using the h.264 encoding {\add resulting in image resolution corresponding to 0.09 mm/pixel}. The resulting exposure time of one frame is 40 ms, which is significantly shorter than the temporal separation between two sequential T1 events \emph{at the same node}. Fig.~\ref{fig1:device}c) displays an example of a
grayscale snapshot overlaid to the schematic top view of the device.
The video is then interpolated to 13~000 grayscale images at 10 Hz frequency {\add, resulting on average 0.1 s time between subsequent frames,} and eventually cropped to a square 
{\add corresponding to the region of interest}. The cropped square can be then used as a grayscale image as is or it can be processed to a skeletonized binary image with exactly one pixel wide lines separating bubbles from each other. The bubbles are thus identified from images as connected 2D regions with an area $A$. Their size distribution using effective radius $R_{eff}=\sqrt{A/\pi}$ is illustrated in Fig.~\ref{fig:bubble_properties} highlighting the monodispersity of our foam. 

In the skeletonized image the nearest neighbors of a bubble are exactly 1 pixel apart. This information is further utilized in identifying the time and location of the T1 events: If there is a change in nearest neighbors, there is either one or more T1 or T2 events. Here, we did not observe any T2 events as no significant coarsening occurs at the time-scales during which the bubbles occupy the imaging area. 
The raw images and the identified locations of the T1 events are then divided into training and test sets as detailed in Supplementary Tables~1~and~2~\cite{supplement}. 

\begin{figure}
    \centering
    \includegraphics[width=\columnwidth]{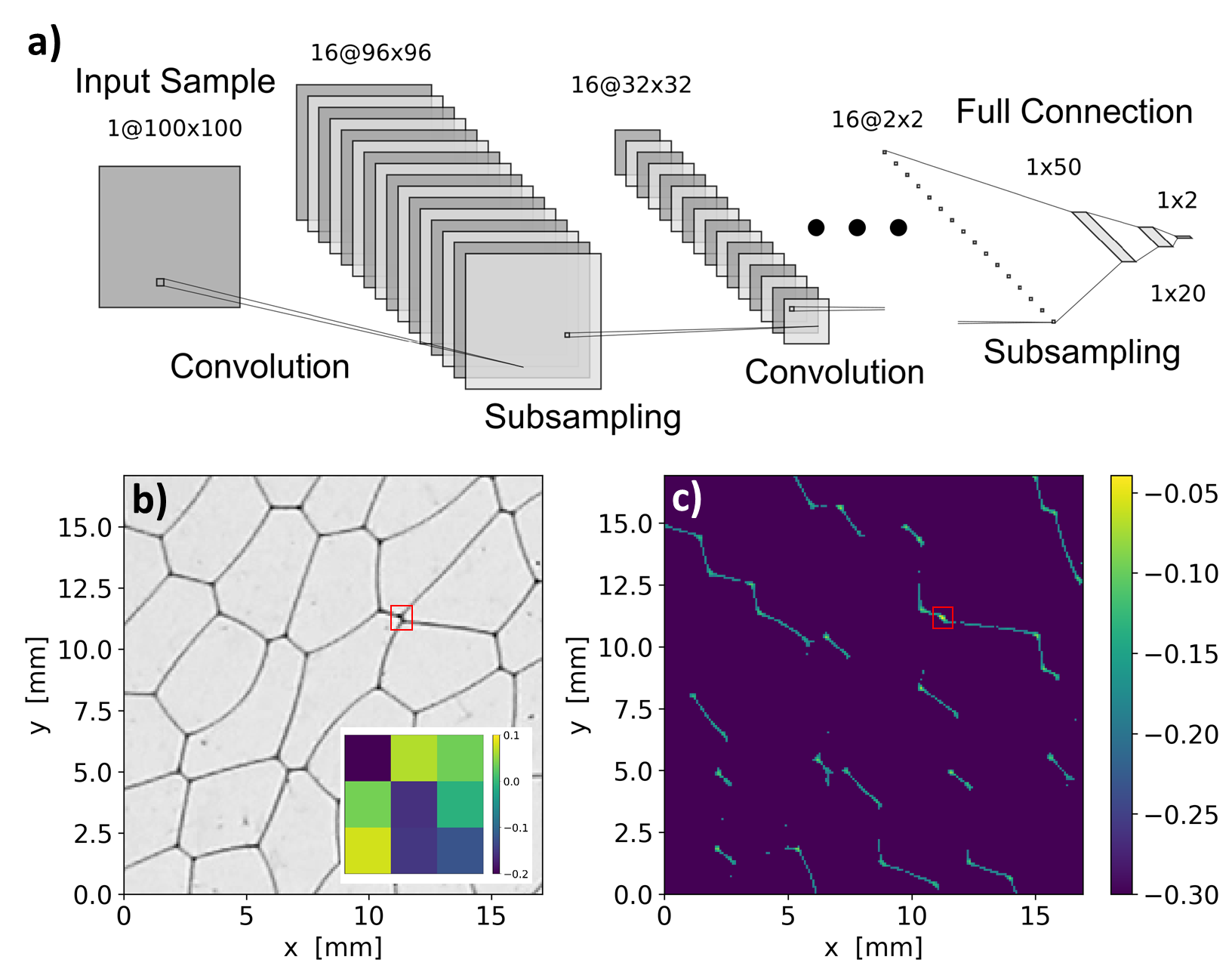}
    \caption{
    a) Architecture of neural network (illustrated using a tool from \cite{nn_picture}) consists of three subsequent convolutional layers and subsampling layers and finishes with two fully connected layers.
    b) Example of a raw grayscale image with one vertex that produces a T1 event highlighted with a red square. The inset shows one of the 3 by 3 convolution filters of a fully trained CNN. The dark diagonal of the filter captures the similarly aligned films of the raw images.
    c) The convolution of the filter and the raw image illustrates the captured elements of the raw image. The convolution is thresholded for clarity.}
    \label{fig:3_ai}
\end{figure}{}

We pose the following question: given a single frame, are we able to predict the occurrence of a T1 event in the subsequent frame or frames?
We start with a convolutional neural network (CNN)~\cite{bishop2006pattern,Salmenjoki2018} and modify it to accept grayscale and skeletonized image as an input.  
Fig.~\ref{fig:3_ai}a) depicts a schematic illustration of the CNN application leading to a single binary classification: either T1 event occurs after $\Delta t$ seconds or not. 
The neural network convolves and coarse grains the input image in three layers. 
The values are then run through fully connected neural network with two hidden layers until one value is left that is thresholded to produce the prediction. 
{\color{comment}In the training, we use Adam optimizer~\cite{kingma2014adam}. In total 10000 training steps are used with learning rate of $5\cdot10^{-5}$, and the prediction results were found to saturate around 2000 training steps. The network was trained using sample batches at each step, rather than using the whole data set at every step. The batch size was 1000 samples and it was tested that increasing the batch size did not improve the prediction results anymore.}
An example of a grayscale input and a trained filter is depicted in Fig.~\ref{fig:3_ai}b) and the first convolution layer is depicted in Fig.~\ref{fig:3_ai}c). Here, the filter is from a fully trained network. Initially the filters contains randomly chosen values that converge to ones capturing essential features by reinforcing the features producing correct prediction using back propagation during training. The other filters capture different orientations of the films, while the function of some filters are not clear.

\begin{figure}
    \centering
    \includegraphics[width=3.4in]{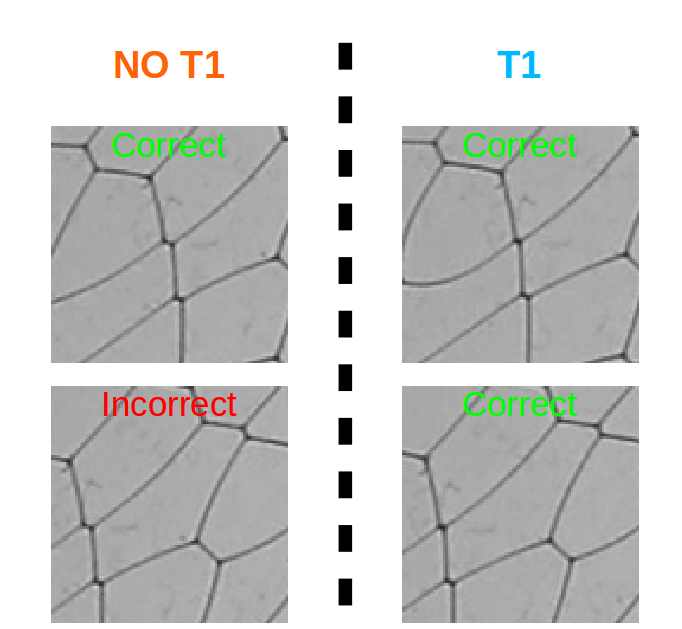}
    \caption{\add Examples of {\color{comment}two} different kind of samples inputted for the CNN and the
    prediction results. On the left are negative samples that do not precede a T1 event within 0.1 
    seconds and on the right are positive samples which precede a T1 event. The sample types are 
    starting from the first row: vertex centered grayscale image and bubble centered grayscale image. 
    The texts ``Correct" and ``Incorrect" in the figure indicate the success of the CNN prediction.}
    \label{fig:ex_samples}
\end{figure}

The input data is split into two, roughly equally large, sets for training and testing.
In the training, a sequence of small regions of images are used.
The essential structural features for T1 detection and prediction may be studied by limiting the amount of input data given to the network, here by limiting the size and location of regions of interest (ROI) around the possible T1 event. 
First, we generate a training set consisting of small ROIs that precede a T1 event in the next frame.
Then, we balance the data set by adding randomly picked non-overlapping regions which do not precede a T1 event until the data set contains 45-55 \% portions of each ROI type.
{\add This results in data sets that contain roughly equal amount of both sample types so the AI does not benefit by favoring either of the outcomes.}
This prescreening is necessary since T1 events are rare. Without it the algorithm scores high by always predicting ``no T1". The quality of the prediction is measured with a score parameter $\xi$ that is the ratio of correct predictions (T1 or no T1) per all predictions. 
{\add Some examples of input images given for CNN are shown in Fig.~\ref{fig:ex_samples} with the outcomes of the prediction.
In Fig.~\ref{fig:ex_samples}, the left column has samples that do not precede a T1 event within 0.1 second time frame while the samples on the right column precede a T1 event.}
{\color{comment}
The figure demonstrates that better accuracy of the predictions can be reached by using the vertex centered sample set than by using the bubble centered sample set. The CNN trained with bubble centered samples highlights features such as bubble shapes and stable vertices that are rather irrelevant for T1 prediction.}

\begin{figure}
    \centering
    \includegraphics[width=\columnwidth]{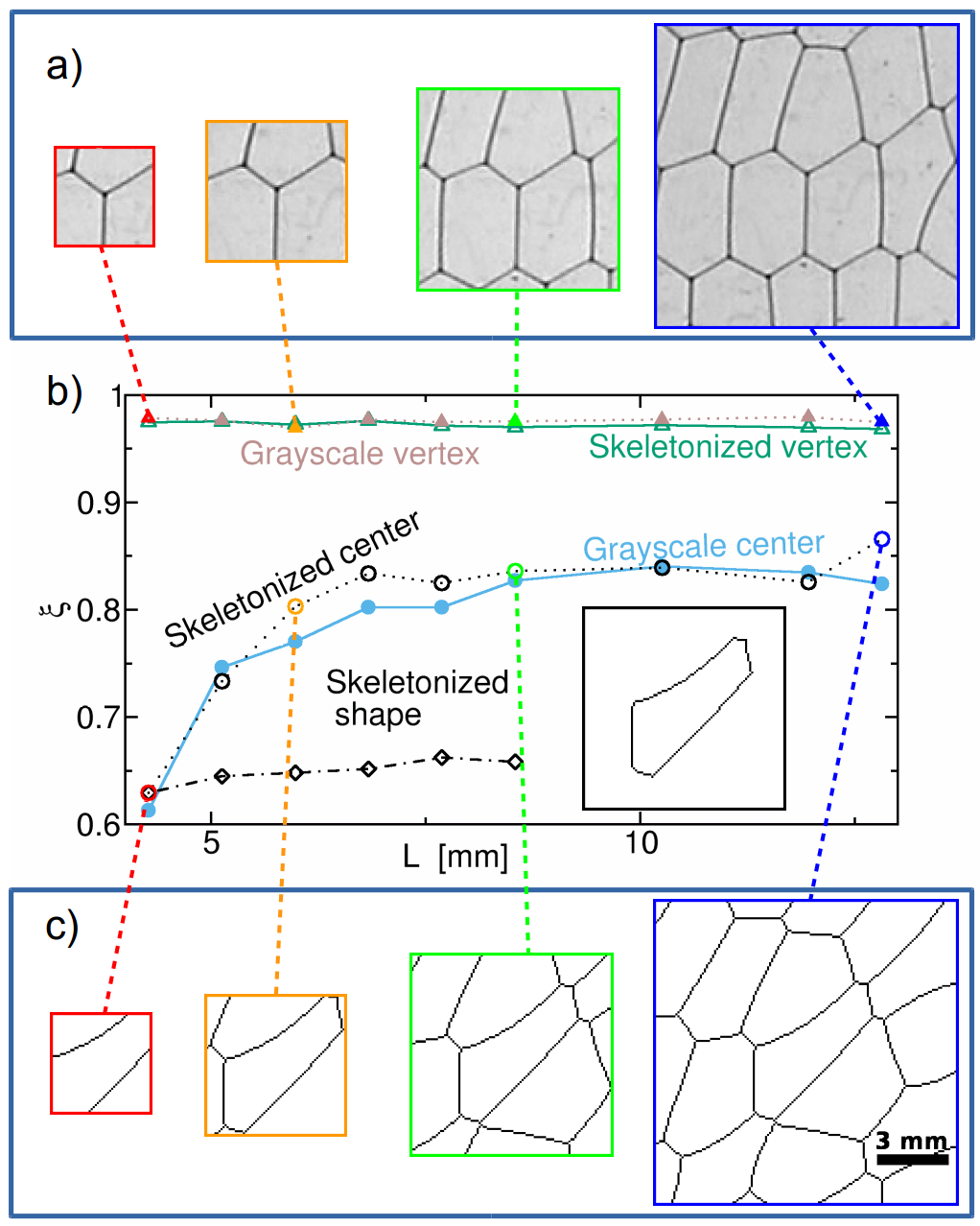}
    \caption{Comparing the score parameter $\xi$ to the size of regions of interest reveals that for vertex centered images the score parameter is virtually independent of the ROI size. However, for bubble centered images
    the accuracy of the algorithm increases
    with the size of the region of interest ($L \times L$ mm) until the area is at least equivalent to that of four bubbles.
    There the performance is similar for both the grayscale (filled symbols) and skeletonized (unfilled symbols) images with the same centering.
    For the largest ROI, the algorithm works best for skeletonized network (unfilled circles) reaching almost 90 \% accuracy. The result is worse for grayscale images (blue)
    or if the data is reduced down to a single bubble shape (inset in b), open diamonds).
 }
    \label{fig3}
\end{figure}

\section{Prediction results}
\subsection{Machine learning}

{\add Next, we explore the T1 predictions produced by the CNN.
For the purpose we use the score parameter defined earlier.}
We {\add study} the CNN for five different combinations of locations and grayscale or skeletonized frames for the full range of ROIs. The combinations {\add selected allows us to} compare i) the effect of film or vertex thicknesses by comparing the predictions {\add produced using} grayscale (filled symbols) and skeletonized images (unfilled) or ii) the effect of {\add focusing on the} vertices (triangles) or bubble centers (circles) or iii) concenrtrating on the shape of the bubble {\add perimeter} without its neighbors (diamonds).
{\add These choices of ROIs allow us to explore the image features, such as bubble shape, liquid fraction, or neighbor bubbles, by which the CNN is able to produce the best predictions.}  
Fig.~\ref{fig3}b) shows the score parameter $\xi$ for different sizes ($4 \leq L \leq 13$ [mm]) and locations (bubble or vertex centered) of ROIs.

The best score $\xi= 98$ \% is obtained using vertex centered ROIs of $4 \times 4$ mm illustrated by the red triangle and a corresponding example input in panel \ref{fig3}a). As a sanity check, we get the same score $\xi_+=98$ \% if we restrict to true positives by defining $\xi_+$ as correctly predicted T1 events per T1 predictions only. This indicates that there is no bias in the classification. 
Increasing the size of the ROI does not improve the score $\xi$ and thus we conclude that the information in the local surroundings of the vertex provides a good indicator of a T1 event (filled triangles).

We considered that the change in the local liquid fraction is the reason for lower yield point and T1 events, what would appear as darker nodes. Thus, removing the information about the local liquid fraction by skeletonizing, we expected that the score parameter would decrease its value significantly. 
We performed the same CNN analysis on the skeletonized frames, which result roughly the same scores with the gray scale images as witnessed in Fig.~\ref{fig3}b) (unfilled triangles).
This is indicative that the CNN is not in fact capturing the gray scale levels in the images,
but rather predicts solely based on the orientations of the films surrounding the vertices.
In other words, local liquid fraction or liquid motion in the films plays no significant role in the local yielding.
Similarly to the gray scale data, the skeletonized data shows no significant dependence on the ROI size.

Moving the ROI center from the vertex to the bubble center of volume has a significant impact on the score parameter.
First, let us concentrate on the skeletonized bubble shape (black diamonds and inset in Fig.~\ref{fig3}b)) all the information of the neighboring bubbles is removed from ROIs.
The score is close to $\xi = 65$ \% with only weak increase with ROI size $L$. 
This is significantly lower compared to the ROIs that focus on the vertex indicating that the bubble shape does not contain enough information {\add for accurate prediction of T1 events}. 
The score calculated only for the positive predictions is even lower $\xi_+ = 60$ \% showing that the CNN does a slightly better job at predicting
samples without a T1 event than predicting the events.

Now we include the films of neighboring bubbles but keep the ROI center at the center of bubble. 
This data is plotted as a function of the bubble centered ROI size for both  grayscale (filled circles) and skeletonized (unfilled circles) images in Fig.~\ref{fig3}b).
The predictions on the smallest ROIs for skeletonized or grayscale images gives similar scores as the predictions based on bubble shapes considered above.
For both skeletonized and grayscale ROIs, the score $\xi$ increases with the size of the ROI. 
As the ROI increases to include the entire bubble and parts of its neighbors (green circle), the score reaches over 80 \%, lower than the one that focused on the vertices, yet
significantly higher than the one excluding all the bubbles neighbors.
This supports our previous interpretation that the essential information on the T1 event is encapsulated by the structures and locations of the vertices and not by the bubble shape.

Comparison of the pure skeletonized data with and without the neighboring bubbles offers a visual confirmation to these observations. These structures are shown in the inset of Fig.~\ref{fig3}b) and highlighted by green color in Fig.~\ref{fig3}c).
In the frame in question, a T1 event will take place in the lowest vertex, a location that is not obvious by only looking at the bubble shape. Thus, if the information on the neighboring bubbles is removed, the score parameter understandably drops dramatically.

\subsection{Comparison to established methods}

\begin{figure}
    \centering
    \includegraphics[width=3.4in]{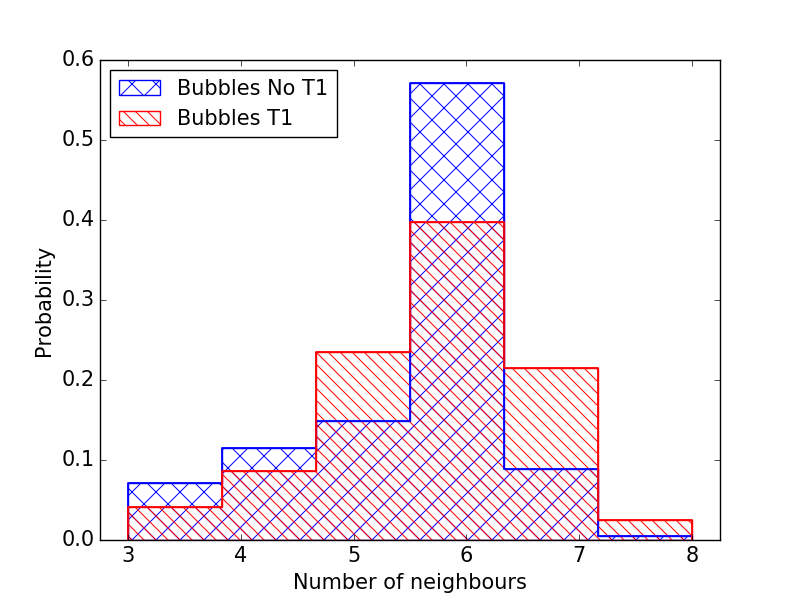}
    \caption{Histogram of number of nearest neighbors for all bubbles and 
    bubbles right before T1 event. Clearly the distribution is wider for the bubbles about to have T1 event indicating that local configurations where bubble has exactly six neighbors are more stable than configurations where bubbles have more or less than six neighbors.}
    \label{fig:neighbor_dist}
\end{figure}

In the simplest view, the CNN could be predicting the T1 events simply by applying 
the Plateau's laws.
{\add This seems to be the case in the sense, that the CNN obtains the best predictions using the vertex centred samples and the Plateau's rules apply to the vertices here. As shown in Fig.~\ref{fig:neighbor_dist} the number of neighbors, on the other hand, only correlates weakly with the probability of the occurrence of a T1 event. As the Figure shows, the most probable number of neighbors for bubble going through a T1 event is actually six. This observation may explain why bubble centred samples yield worse predictability compared to vertex centered.}
Here, {\add however,} one must note that we have a finite accuracy in determining if the Plateau's rules are obeyed. Supplementary Video~\cite{supplement} highlights the lifetimes of these metastable states apparently violating the Plateau's rule lasting up to several seconds. The actual violation of Plateau's rule lasts only a few milliseconds, observed in a 3D case with high magnification \cite{Petit2015}. 
{\add Therefore, more rigorous statement is that the CNN bases the predictions on apparent violation of Plateau's rules within the spatial and temporal resolution of the measurement.}

\begin{figure}
    \centering
    \includegraphics[width=3.4in]{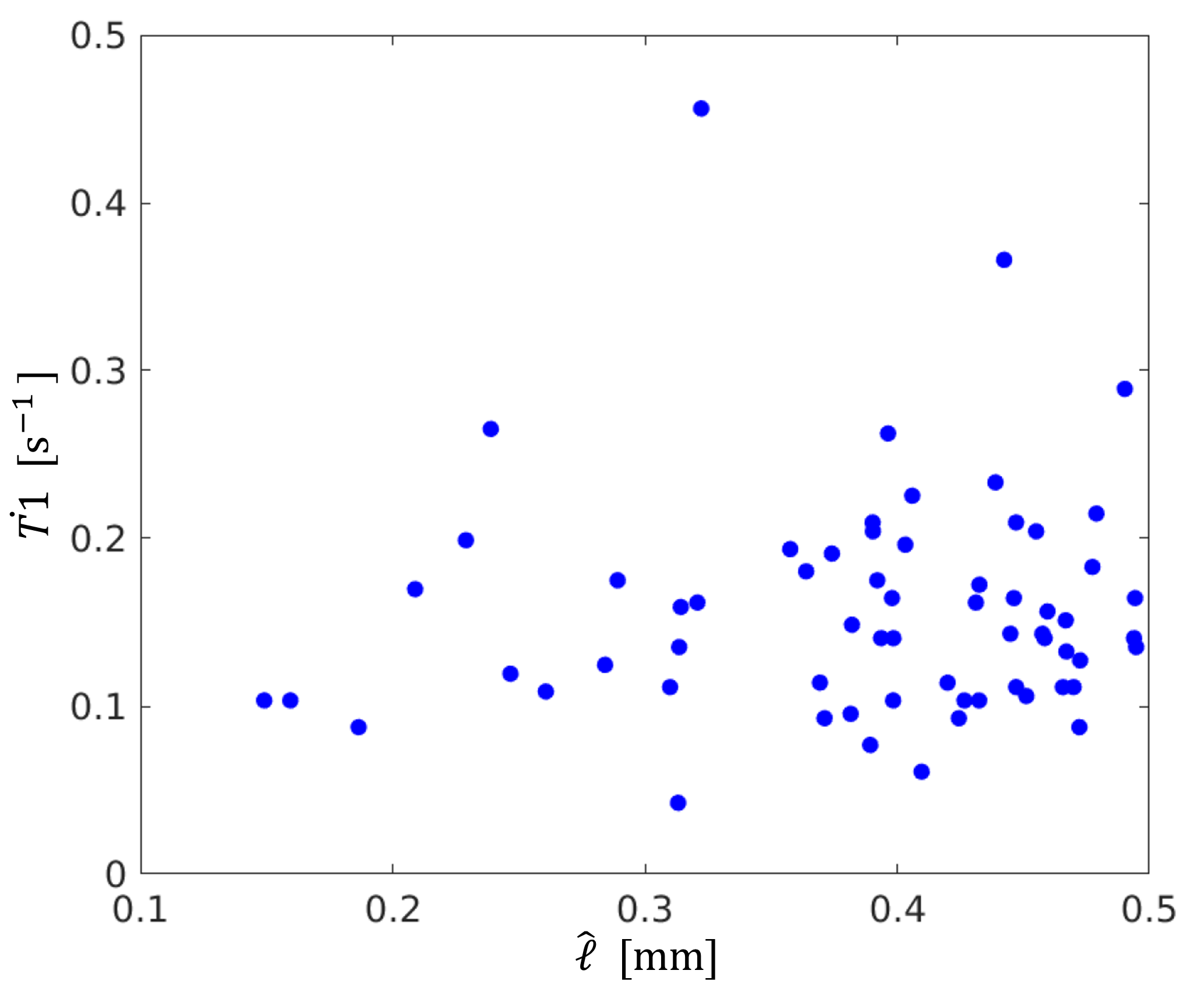}
    \caption{Textural tensor and T1 event rate averaged over $8\times8$ sub-regions of imaged area and time. Although, both quantities show variation between different locations the quantities are not correlated.}
    \label{fig:textural_tensor}
\end{figure}

To further confirm 
the {\add irrelevance of the bubble shape}, we checked for any correlation with the textural tensor 
\begin{equation}
\hat{\ell} = \left\langle \frac{\vec{\ell} \otimes \vec{\ell}}{\ell} \right\rangle
\label{elasticstress}
\end{equation}

\noindent capturing the bubble shape \cite{Marmottant2008} and the T1 event rate which has already been used for various bubble monolayers with similar results \cite{Dollet2007,Dollet2010}. Here $\vec{\ell}$ represents the vector joining two neighboring Plateau borders within a single bubble. 
Using a rectangular mesh of $8 \times 8$ boxes 
{\add dividing} each image {\add into 64 sub-regions}, Fig.~\ref{fig:textural_tensor} shows no correlation between T1 events and the bubble shape for single frames.

Supplementary Tab.~3 shows the score parameter $\xi$ along with other benchmark parameters of our CNN with various sanity checks~\cite{supplement}. These included a virgin data set without any reduction of vertices, comparison to a hand made algorithm 
and prescreening the input data to contain only vertices involving four films, those appearing to violate Plateau's 120 degree rule.
The last method has the best performance, reaching $\xi=99$ \% score and capturing 34 \% of all the T1 events.

Based on our extensive testing we conclude that there exist a significant subset of T1 events triggered by the changes of local microstructure and film orientations in the vertex that can be captured with a CNN without any information on the time dependence.
Thus, already a small region of interest enables the CNN to predict the T1 events.
{\color{comment}In the future, to further improve the predictions we will train other neural networks using parameters, which describe features of the local bubble geometry. We suspect adding some information about the magnitude and direction of the velocity or recent T1 history could improve the predictability.
}
 
\section{Time-dependent properties of elementary yield events}

\begin{figure}
    \centering
    \includegraphics[width=\columnwidth]{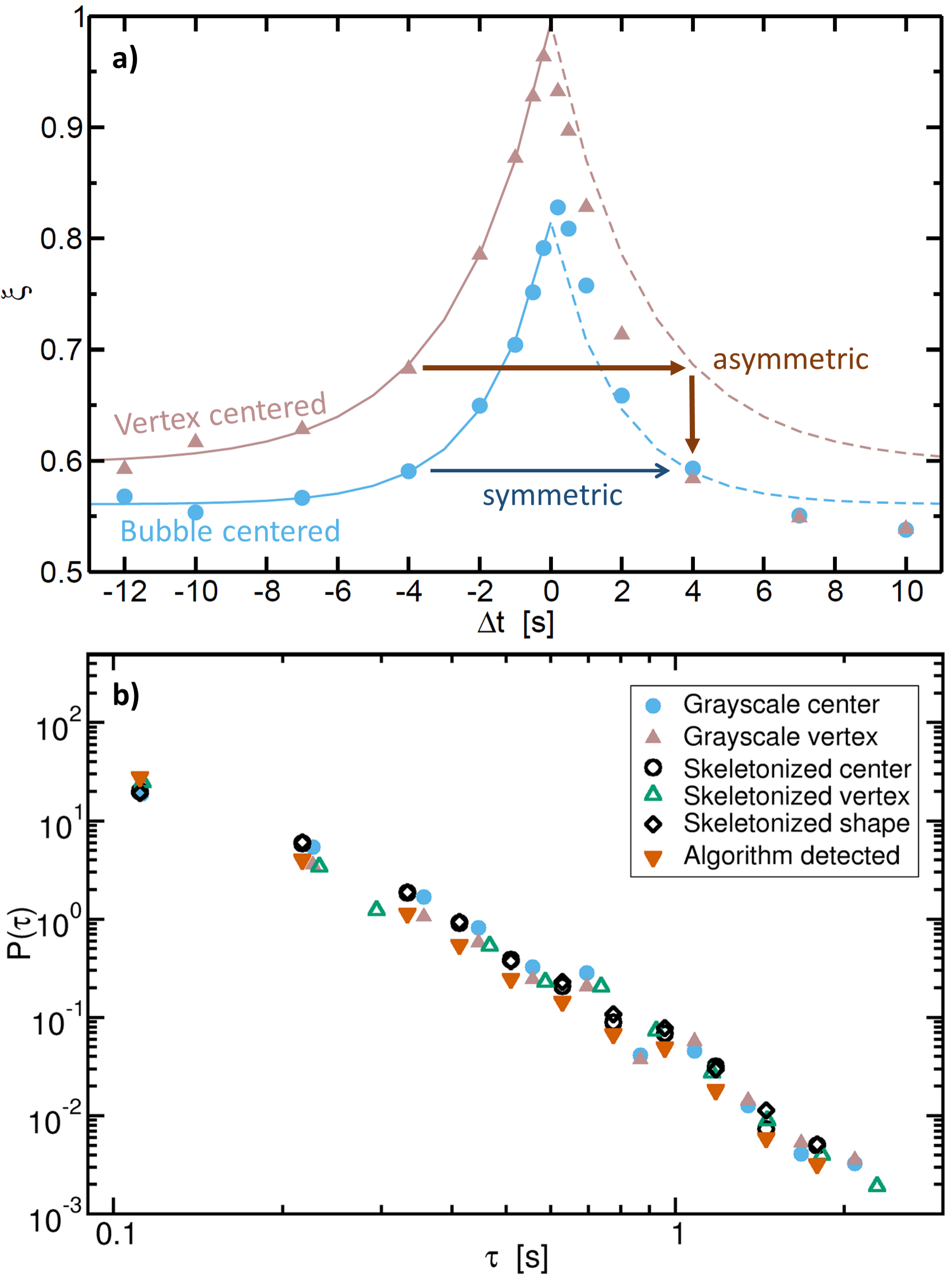}
     \caption{a) Temporal evolution of the score parameter $\xi$  as a measure for structural inhomogeneity for different lag times between the CNN input frame and the T1 event $\Delta t = t_{frame} - t_{event}$, for the vertex (brown triangles) and bubble (blue circles) centered cases. The solid curves are the exponential fits to the data before the T1 event. The same fits are also plotted at  the positive side (dashed). This shows the time reversal asymmetry with the vertex centered case highlighted with arrows. b) The characteristic time scale of subsequent T1 events is of the order of one second from the waiting time distributions $P(\tau)$. 
    } \label{fig4}\label{fig:ksiVsDeltaT}
\end{figure}

We next study to which extent in time the T1 events are predictable.
To answer this, we study the temporal development of T1 events using the data plotted in Fig.~\ref{fig4}.

We identify the score parameter $\xi$ as a measure of predictability.
Essentially, predictability here means the existence of a heterogeneous feature or ``a defect" - since it leads to yielding - different from a featureless material, allowing the CNN algorithm to make a prediction.
We perform the analysis for vertex and bubble centered grayscale samples separately for the ROI size $9 \times 9$ mm$^2$, corresponding to sample image highlighted green in Fig.~\ref{fig3} a) and c).
The size is chosen as the smallest area where the score parameter is saturated to its limiting value.
Also, it is not necessary to include analysis with skeletonized images since the skeletonization only has small effect on the prediction score ($< 5$ \%) if any.

Fig.~\ref{fig:ksiVsDeltaT}a) depicts predictability $\xi$ against the time difference $\Delta t = t_{frame} - t_{event}$ between the T1 event and the single input frame used by the CNN. 
This allows to examine 
the structural signatures of a T1 event before ($\Delta t < 0$), or even after ($\Delta t > 0$) the event. 
The brown triangles refer to data set with vertex centered 
ROI, where the CNN is retrained and evaluated for each measurement point.
Here, the vertex is always at {\add the} center of the ROI even if the lag $\Delta t$ is increased. 

We find that the prediction probability follows an exponential fit (solid line) and drops to one half at $\Delta t = -1.9$ s (before) the event.
Using the ROIs from frames after the T1 event the probability drops to one half faster at $\Delta t = 1.5$ s indicating that there is an asymmetry in predictability.
{\color{comment} Interestingly, a similar breakdown of time reversibility related to local geometry of T1 events has been reported previously in bubble raft shearing experiment measuring orientation of T1 events~\cite{Wang2007}.}
The asymmetry seems reasonable as shear rate drives the film shrinkage while force balance at the liquid-air interfaces drives the film growth for small shear rates (see supplemental video~\cite{supplement})~\cite{biance2009topological}. 
Although, this picture is only crude simplification as evident based on {\add the} video and the previous studies T1 events also show clustering due to redistribution of stress~\cite{desmond2015measurement} causing non-local deformation in the foam~\cite{goyon2008spatial,elias1999foams}.

The bubble centered predictions are symmetric {\add in $\Delta t$}, although the prediction probability is not as good as in the vertex centered case.
This, might be due to {\add the} CNN focusing more on other parts than the film created during {\add a} T1 event. 
Therefore, while the film instability triggers the event, the local neighborhood still plays a significant role.
This can be understood in the context of slow energy dissipation, where a relaxation of the energy landscape enables reversible T1 events \cite{Lundberg2008}. 
Thus, the configuration does not essentially change in the scale of few bubbles, even if the local film geometry can become completely different. 

Fig.~\ref{fig:ksiVsDeltaT}b) shows the T1 event waiting time distributions using the inter-event arrival time in a restricted area 
of $50\times50$~mm.
The data based on all the CNN prediction algorithms agree with that obtained using the nearest neighbor detection algorithm capturing all the events (red triangles).
Moreover, the data shows that the typical T1 time scales are short, of the order of one second in agreement with Fig.~\ref{fig:ksiVsDeltaT}a).
To compare with other dry foams with synthetic surfactants that do not affect the viscosity of the carrier liquid \cite{Denkov2009SM} the time scales for stress jumps was measured 
$t_\mathrm{jump} = 1.5\pm 0.5$ s \cite{Cantat2006} and the average time to reach 90 \% of the final film length was measured $t_\mathrm{90} = 0.5$ s \cite{Durand2006}, that is in both cases of the same order of magnitude as here suggesting that film growth is driven by interfacial tension.
With the average bubble velocity of $v=1.0$ mm/s the timescales are in the ballpark of the average bubble diameter of $d = 4$ mm making T1 events highly localized. 


\section{Conclusions}

We have successfully trained a general purpose CNN to recognize the neighbor swap T1 events in radial 2D foam flow using only snapshots of the structure with no time dependent information.
We capture the essential features of the images, namely vertices and film orientations. 
Using these features we show that typically T1 events initiate from the unstable vertices {\add that appear to violate Plateau's rules} while the bubble shape is a less relevant quantity (Fig.~\ref{fig3}). 
This highlights the importance of local film geometry and microstructure in rearrangements similarly to bubble coalescence \cite{Forel2019}.
The development of the shape or {\add perimeter} of a bubble and its neighborhood is symmetric in time for the bubble participating to a T1 event. In contrast, the changes in films are asymmetric in time.
Here, the emphasis is on the neighbor swap aspect, the bubble is still a relevant unit for other processes such as coarsening \cite{Khakalo2018} and recoverable elastic response \cite{LundbergPRE2008}. The elastic energy stored in the system does not vanish instantly; major part of it is stored in a different film configuration making it possible to have reversible local rearrangements.

On average, a vertex about to yield looks very different before than after the event, as seen for example in Supplementary Video 1, 
that is, one can see whether the configuration under scrutiny is close to the local yield threshold or not~\cite{supplement}.
This manifests as temporal asymmetry in the score $\xi$ 
that drops relatively fast to the baseline after the event.
In other amorphous solids e.g.~granular pillars \cite{CubukPRL2015} the local configuration, namely the particle contacts correlate with the yield stress and exhibit local variations (soft spots). 

Our work has focused on the experimentally amenable 2D case {\color{comment} considering the case of constant driving pressure and liquid fraction}. It would now be quite interesting to investigate how one can change the foam yielding in terms of the local predictability of T1 events and their spatio-temporal correlations. {\color{comment}This could be achieved altering the physical properties of the sample foam, such as liquid fraction, polydispersity, geometry or shear rate. Although, changing any of these parameters may impose difficulties in maintaining a stable flow and keeping coarsening negligible.} One particular direction would be particle-laden foams where the reinforcing particles would induce other dynamical timescales. This leads to a wide variety of industrial applications with tunable orientation dependent properties such as tensile strength or heat conductivity \cite{Wicklein2014, Yang2017}.

\begin{acknowledgments}

JK and AP acknowledge the funding from Academy of Finland (308235 and 278367), Business Finland (211715) and Aalto University (974109903) as well as Aalto Science IT project for computational resources. LV acknowledges the funding from the Vilho, Yrj\"o and Kalle V\"ais\"al\"a Foundation via personal grant and Academy of Finland (278367). {\color{comment}We acknowledge Henri Salmenjoki for sharing CNN code and helpful discussions.}
\end{acknowledgments}


\providecommand{\noopsort}[1]{}\providecommand{\singleletter}[1]{#1}%
%

\end{document}